# Early Biomarkers and Intervention Programs for the Infant Exposed to Prenatal Stress.

Marta C. Antonelli*[1,2#], Martin G. Frasch[3#], Mercedes Rumi[1], Ritika Sharma[2,4], Peter Zimmermann[2], Maria Sol Molinet[2] and Silvia M. Lobmaier[2].

[1]Instituto de Biología Celular y Neurociencia "Prof. E. De Robertis", Facultad de Medicina, UBA, Buenos Aires, Argentina; [2]Department of Obstetrics and Gynecology, Klinikum rechts der Isar, Technical University of Munich, Germany; [3]Department of Obstetrics and Gynecology, University of Washington, Seattle, WA, USA; Center on Human Development and Disability, University of Washington, Seattle, WA, USA; [4]Research Unit of Molecular Epidemiology, Institute of Epidemiology, Helmholtz Zentrum Munich, Neuherberg, Germany.

*# these authors contributed equally to this work*

**Running title:** Biomarkers and programs for at-risk children

**Corresponding author:**

Dr. Marta C. Antonelli. Laboratory of Perinatal Programming of Neurodevelopment, Instituto de Biología Celular y Neurociencias "Prof.E. De Robertis", Facultad de Medicina, Universidad de Buenos Aires, Paraguay 2155 3er piso, 1121 Buenos Aires, Argentina. Phone: +54 11 5285-3266. and Frauenklinik und Poliklinik (Hans Fischer Senior Fellow IAS-TUM). Klinikum rechts der Isar, Ismaninger Str.22. 81675, München, Germany. Phone: +49 89 4140-6671.

E-mail: mca@fmed.uba.ar

**CONFLICT OF INTEREST**

The authors declare no conflict of interest, financial or otherwise.

**ACKNOWLEDGEMENTS**

This study is part of the FELICITy project funded by Klinik und Poliklinik für Frauenheilkunde, TUM, Klinikumrechts der Isar, Munich to SLM and by a Hans Fischer Senior Fellowship from IAS-TUM (Institute for Advanced Study-TUM, Munich), to MCA.

MCA has conceived and designed the concept and road map of the manuscript, drafted it, searched the literature and coordinated the edition of the final version. MGF greatly contributed to the final version of the manuscript by thoroughly reviewing and editing all versions of the manuscript and helping with the design of the illustration. MR prepare and managed the bibliography and review the manuscript. RS and PZ critically review the manuscript. MSM helped in the literature search and critically review the manuscript. SLM helped in the preparation of the manuscript and critically reviewed it. The authors gratefully acknowledge Hermona Soreq lab (Jerusalem, Israel) for collaboration in choline esterase approach. We are greatly indebted to Ezequiel F. Gómez de Lima for producing the illustrations.

**ABSTRACT**

Functional development of affective and reward circuits, cognition and response inhibition later in life exhibits vulnerability periods during gestation and early childhood. Extensive evidence supports the model that exposure to stressors in the gestational period and early postnatal life increases an individual's susceptibility to future impairments of functional development. Recent versions of this model integrate epigenetic mechanisms of the developmental response. Their understanding will guide the future treatment of the associated neuropsychiatric disorders. A combination of non-invasively obtainable physiological signals and epigenetic biomarkers related to the principal systems of the stress response, the Hypothalamic-Pituitary axis (HPA) and the Autonomic Nervous System (ANS), are emerging as the key predictors of neurodevelopmental outcomes. Such electrophysiological and epigenetic biomarkers can prove to timely identify children benefiting most from early intervention programs. Such programs should ameliorate future disorders in otherwise apparently healthy children. The recently developed Early Family-Centered Intervention Programs aim to influence the care and stimuli provided daily by the family and improving parent/child attachment, a key element for healthy socio-emotional adult life. Although frequently underestimated, such biomarker-guided early intervention strategy represents a crucial first step in the prevention of future neuropsychiatric problems and in reducing their personal and societal impact.

# 1.  GESTATIONAL ENVIRONMENT AND EARLY PARENTING

The spread and depth of evidence from both animal and human studies leave nowadays little doubt about the impact of the gestational environment on the fetal brain development; a concept that has been named "fetal programming" [1, 2]. Prenatal stress (PS) impacts early behavioural, cognitive development and temperament in human infants and increases child morbidity and neurological dysfunction such as attention-deficit hyperactivity disorder (ADHD) and sleep disturbance during infancy which if persistent in adulthood might result in depression and vulnerability to psychotic disorders [3, 4]. In their review, Van Den Bergh *et al.* [3] concluded that numerous epidemiological and case-control studies of the past decade show neurodevelopmental disorders in offspring exposed to maternal stress during pregnancy. Reviewed studies mainly refer to pregnant women exposed to psychosocial stress including states of anxiety, depressive symptoms, major life events experienced by the mother, experience of a disaster and subjective distress during the third trimester of pregnancy. These pregnant women have infants that show: less affective reactivity at 5 months [5], higher temperamental reactivity at 6 months [6], positive association with high respiratory sinus arrhythmia at 8-10 months [7], higher negative affectivity at 24 months [8], higher reaction intensity at 24-30 months [9], and decreased cognitive functions at the age of 24 months [10]. Together, these findings point to the main outcome of the PS exposure being at the cognitive and emotional regulatory levels. More recently, Persson and Rossin-Slater [11] showed that *in utero* exposure to bereavement has long-lasting effects on the consumption of psychiatric medication both during childhood and adulthood. Children are 25% more prone to taking medication to treat ADHD and experience a 13% and 8% increase in the consumption of prescription drugs for anxiety and depression, respectively, when they reach adulthood. These data highlight the serious economic burden of pregnancy-associated mental illnesses generating important private and societal costs. For example, in the United States the market for antidepressant drugs totaled $9.6 billion in 2008. The consumption of prescription drugs for treating ADHD increased five times in a decade from $1.7 billion in 2003 to $9 billion in 2013. Persson and Rossin-Slater [11] also pointed out that in Sweden, "*mental illness accounts for a larger share of health expenditures on prescription drugs that any other therapeutic drug*". Despite these insights, we observe that the long-term health–economic impact of pregnancy-associated psychiatric illness remains highly underappreciated, especially considering that early interventions during pregnancy and/or nursing periods can recover these high costs to individuals and society.

It has also been acknowledged that not only intrauterine adversities due to maternal stress influence the development of the child, but also a low maternal involvement during upbringing will influence the child´s neurodevelopmental outcome [9, 12-15]. Several studies have shown that high levels of pregnancy-specific anxiety and trait anxiety during pregnancy may persist postnatally predicting higher levels of parenting stress three months after birth and leading to a lower parenting self-efficacy and negative perceptions of parenting-related issues [16] suggesting that anxiety during pregnancy interferes with an optimal preparation for parenting. This means that maternal stress during pregnancy and during early parenting can program physiological responses and lifetime trajectories of the infant, which in interaction with genetic liabilities and early-life challenges, will determine the ultimate health status. The concepts derived from these studies contributed to the emphasis on maternal health as a global priority for the World Health Organization (http://www.who.int/pmnch) and the International Monetary Fund (http://www.imf.org/external/np/exr/facts/mdg.htm). WHO (2016) states that: "*The burden of mental disorders continues to grow with significant impacts on health and major social, human rights and economic consequences in all countries of the world.*" This reinforces the idea that nurturing care in early life is essential to enable children to become healthy and productive citizens with adequate intellectual skills, creativity and wellbeing [17, 18]. This means that investing in childhood as early as possible will impact the development of a healthy society.

## 1.1 Developmental Hypothesis

Several developmental models have been reported based on Barker's Fetal Basis of Adult Diseases (FeBAD) hypothesis. Since 1987 and based on their studies on adult cardiovascular diseases, Barker and his group elaborated the hypothesis that the *in utero* environment may permanently program the structure and physiology of the offspring [19, 20]. The original FeBAD hypothesis evolved to include adaptive responses or phenotypic plasticity [21]. The term FeBAD was later changed to include the term Health, introducing the notion that not only diseases can be shaped during the perinatal period but also the health outcomes. The name of the hypothesis was then agreed to become the Developmental Origins of Health and Disease (DOHaD) [22, 23]. An extension of this hypothesis was recently developed by Van den Bergh *et al* [24] who proposes the "Developmental Origins of Behavior, Health and Disease" (DOBHaD) hypothesis, introducing two fundamental concepts: a) specific signs may announce disorders to come before they appear and 2) in spite of the negative impact of perinatal adversities, developmental plasticity allows timely changes for reversion. The DOBHaD hypothesis integrates concepts from the epigenetic field and reinforces the idea that understanding the process by which development responds to an insult can guide the future treatment of the disorder.

## 1.2 Biological mechanisms

In recent years, much effort has been devoted to understanding the biological mechanisms that underlie the basis of neurodevelopmental disorders triggered by PS. Two questions arise sequentially: 1) What are the potential mediators that connect the stressed mother with the fetus? and 2) How do these mediators leave "permanent" signatures in the baby? The first question has been thoroughly reviewed by Rakers *et al* [4] and they conclude that the main mediators of maternal-fetal stress are not only the well-known cortisol but also catecholamines, reactive oxygen species (ROS), cytokines, serotonin/tryptophan and maternal microbiota. This implies that not only the re-programming of the Hypothalamic-Pituitary axis (HPA) is involved but the Sympathetic-Adrenal Medullary System (SAMS) as well, among other systems.

The second question is still a matter of intense study but there is a general consensus that epigenetic mechanisms are the most probable link between prenatal environmental exposures and the disruption of normal brain function [25, 26]. Most of these mechanisms and signals have been proposed as predictive biomarkers of neurobehavioral outcomes but in spite of much promising research efforts, we still lack well-designed longitudinal studies of maternal/fetal dyads with or without stress exposure that will definitively and unequivocally correlate biological signals with the child behavioral outcome. These biological signals might help to predict adverse outcomes and to involve the child in early stimulation programs.

The importance of finding reliable early predictive biomarkers is highlighted by the fact that the stress situation is often not detected on time or the pregnant/nursing mother is unaware of the potential effect of stress on the disadvantageous outcome for her baby. It is well known that the brain is particularly sensitive to changes in the perinatal environment during early development, but the consequences of prenatal damage may not necessarily be apparent until a critical age when neurodevelopmental defects may be precipitated by a subsequent exposure to other insults. Most frequently, negative outcomes in children are discovered many years later missing many opportunities and years of adequate stimulation programs. In fact, several studies reported that various developmental disorders were detected at school age in apparently healthy children. Screening trials in healthy children from 0-5 years old showed a 16-20% prevalence of global developmental and communication disorders [27-29].

As stated by Rakers *et al*. [4], we believe that the effects of maternal stress on fetal development are mediated by a *"multiple stress-transfer mechanisms acting together in a synergistic manner"*. We propose that a combination of multimodal biomarkers will help to detect "at-risk children" as early as possible in order to make the decision to face early stimulation programs, whenever no prevention measures in the mother can be timely taken [30]. In this review, we will describe some of the combinations of signals and biomarkers as well as early intervention programs that have proven effective in reversing or ameliorating neurodevelopmental disorders in children exposed to prenatal stress.

## 2.    BIOMARKERS FOR EARLY DETECTION

The importance of early detection of children at risk of future neurodevelopmental sequelae has been clearly delineated by Braun *et al* [31]. Although much work is still needed, there are clear hints from animal and humans studies that after controlling for time of stress exposure, early interventions in children can ameliorate, "reverse" or "repair" the cognitive deficits induced by PS. As mentioned before, when the mother is unaware of her stress or the newborn has no apparent behavioral symptoms, the availability of predictive biomarkers is urgently needed.

Gene-environment interactions generally involve epigenetic changes and these changes can leave stable marks in the genome in response to the environment, potentially altering the gene expression for life and even transgenerationally [32]. Indeed, this phenotypic stability and reversal under intervention are features that give these epigenetic changes the potential to become predictive biomarkers that allow early intervention and prevention of neurobehavioral risk. This hypothesis has been proposed and developed in numerous studies both in humans and in animal models and most of them found several genes with alterations in the methylation profile that could be identified as biomarkers. However, we believe these studies have two important limitations: a) most of the studies were carried out using candidate genes interrogating the HPA axis, on the basis that stress alters this axis, disregarding the multigenic nature of the stress response genes; and b) few studies have assessed altered sympathetic and vagal activity during fetal life as a possible consequence of prenatal stress. In the following paragraphs we will describe the most studied potential biomarkers, the epigenetic signatures and the emerging candidate biomarkers related to the Autonomic Nervous System (ANS). We refer to ANS as including SAMS and the parasympathetic (i.e., mostly vagus nerve) activities.

### 2.1. Epigenetic biomarkers

The process of 'fetal programming" is mediated by the impact of prenatal experience on the developing HPA axis. The HPA is a dynamic metabolic system that regulates homeostatic mechanisms, including the ability to respond to stressors [33] and which is highly sensitive to adverse early life experiences [34]. It has been suggested that the latency between the exposure to stress and the occurrence of the disease is pointing out to the fact that the environment triggers stable changes that have the potential to manifest later in life [35].

Several recent reviews support the hypothesis that fetal programming is mediated by epigenetic mechanisms that persistently alter gene transcription affecting physiology and behavior [25, 36]. Epigenetic mechanisms change the gene activity or expression altering the chromatin organization without modifying the genetic code. Several processes have been described that stably alter the gene accessibility to the transcriptional machinery such as DNA methylation/hydroxymethylation, histone modifications (acetylation, methylation, ubiquitination and sumoylation) and microRNAs modifications. However, the most highly studied and best characterized epigenetic mark, DNA methylation, involves a direct covalent, chemical modification of a cytosine base lying sequentially adjacent to a guanine base (thus a CpG dinucleotide); such methylation is a relatively stable epigenetic tag, catalyzed by a group of enzymes called DNA methyltransferases (DNMTs) [37]. CpG dinucleotides are relatively infrequent in the genome, and areas of comparatively high CpG content have been termed 'CpG islands'. CpG islands tend to be hypomethylated compared to other CpG sites and are found associated with approximately 70% of known gene promoters, i.e., the regulatory, non-coding portion of a gene that plays a role in transcription control. Promoter DNA methylation (often in CpG islands) and gene body DNA methylation generally show opposite associations with gene expression. The presence of 5-methylcytosine is usually associated with the transcriptional silencing of the underlying DNA sequence [38, 39]. The first evidence for an epigenetic mechanism mediating exposure to a perinatal insult was provided by examining the epigenetic effects of differences in maternal care in the rat [40]. These authors and others found that the offspring of mothers who exhibited reduced litter care (low licking and grooming) present in adulthood increased methylation in exon $1_7$ of NR3C1 (Nuclear Receptor Subfamily 3 Group C Member 1, that encodes the glucocorticoid receptor), a change associated with a decreased expression of glucocorticoid receptor (GR) in the hippocampus and with an exacerbated response to stress [40-42]. Moreover, epigenetic variations have been reported, mostly changes in DNA methylation in several brain regions, in animal models after prenatal exposure to stress [43], maternal separation [44] and response to variations in the mother-offspring interaction [45, 46]. In humans, higher levels of methylation were detected in exon 1F of NR3C1 (homologous exon $1_7$ in rat) in the hippocampus of suicide victims who were, in turn, victims of child abuse [47], in umbilical cord blood cells of children whose mothers were diagnosed with depression and anxiety during pregnancy [48] and in pregnant mothers undergoing chronic and war-related stress [49].

Biomarkers are molecules that characterize the particular signature of a physiological/pathological process and that can be easily accessed and quantified. To evaluate PS outcomes in newborns, biomarkers are particularly needed to reveal epigenetic reprogramming that occurs in the brain, an inaccessible organ in humans [50]. DNA methylation in blood, saliva or tissues such as placenta is, therefore, a promising biomarker candidate since: a) it is easily detectable and quantifiable; b) it is chemically stable and not affected by cyclic fluctuations as cortisol; c) it is an early response mechanism, making it a sensitive indicator. At present, changes in DNA methylation of certain genes can be monitored accurately in samples of human placenta and umbilical cord blood [51]. Alternatively, saliva and buccal epithelium cells represent an additional easily accessible source of DNA and provide information on the systemic condition of an individual. The procedures for obtaining saliva have the advantage of being simple, inexpensive and less invasive since they do not involve any side effects such as bruising, infections, etc. In addition, sampling can be done from early childhood onwards [48, 51]. In a recent study, Essex et al, [52] examined buccal epithelial cells in a cohort of adolescents and found differences in DNA methylation in those adolescents whose parents reported experiencing stress and depression during the adolescent childhood.

The methods for analyzing gene-specific DNA methylation can be divided into **"candidate gene"** and **"genome-wide"** studies. The **"candidate gene"** approach directly tests the effects of genetic variants of a potentially contributing gene in an association study [53] thus excluding a large number of genes. In contrast, the **Epigenome-Wide Association Studies (EWAS)** approach allows an unbiased analysis of locus-specific methylation across an entire genome. Thanks to the incorporation of microarray-based methylation profiling platforms, this technique is now relatively inexpensive. All studies summarized in the preceding paragraphs were performed interrogating candidate genes based on the hypothesis that DNA methylation in brain regions of the fetus is related with the fetal HPA axis response to maternal stress.

However, the multigenic nature of the stress response and neuropsychiatric disorders, is manifested through small and simultaneous changes in the expression of several genes [54]. Accordingly and beyond the HPA axis genes, Vidal et al [55], found alterations in the methylation in the differentially methylated region (DMR) of MEST gene (Mesoderm Specific Transcript) and more recently, Vangeel et al [56], found an association between methylation of a gene IGF2 DMR (Insulin family of Growth Factors) with pregnancy-associated anxiety. However, recent publications employing the EWAS approach have shown contradictory results. Combining data from two independent population-based samples in an EWAS meta-analysis (n=1740 dyads), Rijlaarsdam et al. [57], showed no large effects of prenatal maternal stress exposure on neonatal cord blood DNA methylation. More recently, Wikenius et al. [58] studying prenatal maternal stress (n=184 dyads), in the form of maternal depressive symptoms, found no significant genome-wide association between maternal depressive symptoms and infant DNA methylation. In contrast, Non et al. [59] reported the identification of 42 CpG sites with significantly different cord blood DNA methylation levels in neonates (n=36) exposed to non-medicated depression or anxiety relative to controls. Interestingly, they report that after a gene ontology analysis they found a significant clustering of genes related to transcription, translation, and cell division processes, but no genes were related to the HPA axis. Employing a cross-tissues/cross-species study of the effects of early life stress (ELS) using a genome-wide approach, Nieratschker et al. [60] found that *MORC1* (MORC family CW-type zinc finger 1, a protein required for spermatogenesis) was differentially methylated in humans (n=180 dyads), monkey and rat both in peripheral tissues as well as in the brain and at different time points throughout life-span. Other studies investigated extreme conditions of ELS, such as war situations or natural disasters and found broad effects of ELS on methylation in several genes of the HPA axis in a war-related stress situation (n=24 dyads) [49], while Cao-Lei *et al.* [61] reported that PS in the form of a natural disaster (Ice-Storm in Quebec) was correlated with DNA methylation (in T-cells of 13 years old children) in 1675 CGs associated with 957 genes related to immune function. It is probably too early to draw general conclusions based on these studies since many differences are still observed in terms of the type and timing of prenatal insult, tissue employed and the children´s age.

Histone and miRNA modifications are also promising biomarkers of PS. Even though most of these studies come from animal models, several biomarkers were shown to be related to neurological diseases in humans [62-65]. In our studies, PS was shown to increase miRNA-133 in prefrontal cortex and hippocampus of in PS male rats [66] and an increase in histone methyltransferase *suv39h1* in hippocampus of PS offspring, with no changes in histone deacetylases *hdac2* and *hdac3* mRNA levels [67]. Zucchi *et al.* [65] showed that several miRNA profiles were up-regulated in PS offspring brains (miR145 and miR103) and several were down-regulated (miR323, miR98 and miR219). At the protein level and employing a mouse model, Benoit *et al.* [68] showed that adult offspring exposed to unpredictable chronic stress during pregnancy exhibited decreased acetylated histone H3 (AcH3) in the hippocampus with sex-specific changes. In accordance with the gender differences for miRNA-133 in male rats, Van Den Hove *et al.* [69] showed an up regulation of histone deacetylase 4 protein (HDAC4) in the hippocampus of prenatal restraint stress (PRS) male offspring, while it was down regulated in frontal cortex of PS female offspring.

In summary and in spite of the disparity of these findings, overall there is ample evidence supporting the hypothesis that PS produces stable and long-term phenotypic changes in the offspring involving persistent alterations in gene function through changes in DNA methylation, histone and miRNA modifications. Although further use of the epigenome-wide approach will clarify this association, it seems plausible to envision epigenetic marks as biomarkers for children at-risk of suffering developmental diseases.

## 2.2.    Biomarkers of the Autonomic Nervous System

HPA-axis oriented studies have overlooked other physiological signs. For example, corticosteroid administration during pregnancy has shown to affect autonomic balance in utero [70-73]. This effect is transient but repeated fetal administration of betamethasone alters nervous system maturation. In fact, Braithwaite *et al.* [74] found no association between maternal cortisol and infant DNA methylation suggesting that the effects of maternal depression may not be mediated directly by glucocorticoids; instead, sympathetic nervous system activity, a component of the fetal ANS, may be the mediating pathway.

The fetal ANS has proven to be very sensitive to maternal stress [75-77]. Among other authors, Kinsella and Monk [78] and Gao *et al.* [79] have indicated that common biomarkers of ANS such as Fetal Heart Rate (FHR) reactivity to a stimulus, or heart rate variability, reflect emerging individual differences in the development of the autonomic and central nervous systems related to styles of future emotional regulation and the risk for psychopathology.

 Advanced analysis of FHR patterns specifically assessing changes in the autonomic regulation of FHR may identify the fetus at increased risk for fetal programming. This can be assessed by a relatively new promising method, phase-rectified signal averaging (PRSA) analysis of FHR, measured by electrocardiography (ECG) or cardiotocography (CTG) [80, 81]. PRSA can eliminate signal artifacts and extract areas of interest from FHR. In contrast to other methods analyzing FHR variability, PRSA permits the detection of quasi-periodicities in non-stationary data, an important benefit of the PRSA approach in dealing with an often-noisy FHR signal. Schmidt and his team already reported the excellent performance of PRSA in adult cardiology for predicting mortality after myocardial infarction [80]. This method was first described for fetuses by our own research group [82-84]. PRSA has then been successfully applied in fetal medicine also by other teams, despite the challenges of a non-stationary signal, with more intrinsic disturbances in the signal than in the adult after a myocardial infarction. The novel parameter referred to as cardiac average acceleration and deceleration capacity is more specific than the conventional FHR analyses in identifying intrauterine growth restriction (IUGR) antepartum [83-85], predicting IUGR outcome and strongly correlates with acid-base biomarkers during acute hypoxic stress in humans during labor [86, 87] and the fetal sheep model of labor [88]. Newer data also show an activation of ANS in fetuses affected by maternal gestational diabetes which could not be seen using conventional techniques [89]. To evaluate the ANS influence on FHR using PRSA approach, the beat-to-beat information (R-R intervals of ECG) should be analyzed, although the conventional Doppler CTG signal can also be used for the analysis. The new generation of the trans-abdominal fetal ECG monitors (such as Monica AN24, Monica Healthcare, Nottingham, UK or Invu by Nuvo-Group, Tel Aviv, Israel) allows for a completely non-invasive and passive recording of fetal and maternal ECG: it only records electrophysiological signals from the women's abdomen without hampering mobility or other diagnostic procedures. This signal can then be used for a more sophisticated analysis of FHR such as PRSA or multidimensional FHR variability analysis [90].

We have recently assessed couplings between maternal heart rate (MHR) and FHR as a new biomarker for PS based on a signal-processing algorithm termed bivariate PRSA (bPRSA) yielding fetal stress index (FSI) values by jointly assessing changes in MHR and FHR. In a prospective case-control study matched for maternal age, parity and gestational age, we used the Cohen Perceived Stress Scale-10 (PSS-10) questionnaire to categorize women in the 3rd trimester of pregnancy as stress group and control group. We could detect periodic MHR decreases reflecting typical pattern of maternal breathing (sinus bradycardia during expiration). Interestingly, control group fetuses remained "stable" during these periods whereas fetuses of stressed mothers showed significant decreases of FHR. The proposed FSI provides unique insights into the relationship between two connected biological systems: mother and fetus. It is hence likely that the FHR response to MHR changes represents a fetal stress memory and may serve as a novel biomarker to detect PS effects early in utero which may help guide early interventions prenatally and postnatally [91].

After birth, ANS activity can also be assessed by measuring the baroreceptor reflex sensitivity (BRS). To improve the signal to noise ratio and hence the quality of BRS estimation, non-invasive continuous blood pressure monitoring with Finapres is used to obtain systolic blood pressure information, combined with a simultaneous ECG recording to extract a high-quality R-R signal and compute the pulse. A bivariate PRSA method, akin to FSI, is then applied for BRS calculation [92]. This method allows the correlation of blood pressure with ECG giving complementary insights into ANS. This approach has never been deployed in neonates and infants affected by fetal programming before and seems to be a very promising tool for this patient group.

Human studies pose various challenges in terms of confounding factors and accessibility to brain tissues. Future studies will hence require a combination of non-invasive physiological measures of the stress response system that are unequivocally linked to PS and that could be deployed as predictive biomarkers of the child neurodevelopmental outcomes. By integrating multiple non-invasively obtainable sources of information this framework could yield progress in maternal, fetal and child health, offering a more precise and personalized prediction and new possibilities for designing interventions to improve neurodevelopmental outcomes of pregnancy affected by PS.

## 3.    EARLY INTERVENTION PROGRAMS IN CHILDREN

The biological rationale for designing Early Intervention Programs (EIP) is based on the well-known property of the brain known as cerebral plasticity. As elegantly synthesized by Brenhouse and Andersen [93]: *"While genes provide the blueprint to construct the brain, experience sculpts that brain to match the needs of the environment"*. Assuming that the conditions are favorable, there is a time window during which brain functions develop normally; however, during anomalous conditions, there is a sensitive period during which the structure or function can be modified. This adaptation to the needs of the environment implies that environmental influences such as stress have a significant impact on brain development. One of the roles of cerebral plasticity is correcting anomalous cognitive development and may have favorable effects even if the change it induces does not reproduce the normal sequence of normal development. Since the young brain has a higher potential towards plasticity, intervention programs are devised as early as possible aiming at stimulating the brain during the periods of its plasticity so that such neuroplasticity processes may correct the abnormal cognitive development [94-96].

Although it seems logical to assume that maternal stress during pregnancy would lead to parenting stress during nursing, few studies have addressed this important topic. Among these studies, Huizink *et al.* [16] observed that pregnancy-specific anxiety and trait anxiety predict higher levels of parenting stress 3 months after birth. As observed by Mc Quillan *et al.* [97], postnatal stress leads to poor and insufficient sleep which in turns leads to to less observed positive parenting practices and more self-reported dysfunctional parenting. More recently, Hagaman *et al* [98] showed that, in a peri-urban cohort in Pakistan, depression and stress during pregnancy were significantly associated with poor maternal functioning in everyday life during the first year after birth. It is, therefore, plausible to predict that the neurodevelopmental outcome in terms of affective and reward circuits, cognition, mood, emotional regulation, personality and learning abilities of the infant would be shaped by the antenatal exposure to maternal stress and subsequent less than optimal parenting. Inversely, it has been demonstrated that engaging in positive parenting practices, such as reading to children, engaging in storytelling or singing and eating meals together as a family may influence a child's future success in the educational system and reduce the risk for delays at the developmental, social and behavioral levels [99]. According to Garg *et al.* [100], "*The quality of the early care environment can buffer some of the negative effects of prenatal adversity on child development*" and at the molecular level. These authors demonstrated that the early care environment co-varies with variation in genome-wide DNA methylation in middle childhood meaning that there is an association between infant attachment at 36 months of age and DNA methylome variation in later childhood.

Excluding pathological outcomes, there is again little research directed at correlating the influence of both prenatal and postnatal unfavorable environments on the offspring´s behavior, particularly on how the environment will impact on the infant personality and character. Two recent studies show that environmental harshness and hawk character during the first two years of life predicted worse visual-problem solving at 4 years of age [101] and that children´s temperamental predispositions, paired with a history of regulatory problems in infancy and maternal depressive symptoms contribute to an increased risk of behavioral problems [15], pinpointing the influence of early disadvantageous environments and character on behavioral and cognitive abilities in later life.

In the face of this body of evidence, the development of EIP is urgently needed to ameliorate these undesired behavioral and cognitive outcomes and to aid parents whose babies were exposed to PS. The history of "Early Stimulation Programs", as were originally called, dates back to 1965 when a publicly funded program was created in the United States for vulnerable children in low-income families [102]. Since then, most of the early stimulation programs have been targeted to neurodevelopmentally impaired or preterm or low-birth-weight infants. Even though there are no specific programs designed for babies that were exposed to PS, the wealth of programs developed to stimulate children and to improve parenting abilities are extremely valuable antecedents that will help to choose the appropriate intervention strategy for an early stimulation program oriented toward prenatally stressed babies. As mentioned above, these babies develop as "apparently healthy children" until global developmental and communication disorders are diagnosed many years later, sometimes at school age.

Since summarizing the existing literature on the EIP is out of the scope of this review, we will briefly outline the main benefits reported in selected studies. According to Bonnier [95], EIP have been developed for three different target populations: 1) Low socioeconomic status children at-risk, 2) Children with disorders that induce developmental delays (e.g. Down Syndrome, Cerebral Palsy) and 3) Preterm or low weight babies. Cognition improvements are greater than motor skills [95] although other studies point out that there are moderate effects on cognitive and behavioral outcomes [103] and improved visual and motor skills during infancy [104] but that cognitive benefits persist into preschool age [105]. In spite of the different targeted populations and the heterogeneity between studies, most reviews concur that EIP produce the greatest positive effects when the programs involve both parents and the child and that long-term stimulation programs improved child-parent interactions [95, 96].

In this sense, a recent review [106] and a meta-analysis [107] report the results of several studies employing different programs involving parents and children known as: *Early Interventions focused on the family* or *Early Family-Centered Interventions*. This approach aims at improving the parent-infant interactions and relationship resulting in an improvement in the child´s development and increased resilience. Although its effectiveness has yet to be fully proven, parental psychosocial support is a sensible approach to increase parent´s ability and knowledge to care for their child. Parent´s support is intended to reduce stress, anxiety and depression as well as improving the parent´s attunement and capacity to interact with their child, which will eventually improve the developmental outcomes. Intuitively, it is clear that building a strong parent-infant relationship will help the infant learn self-regulatory skills such as crying but being consolable, self-soothing, sleep-wake changes and feeding [106]. As mentioned before, most of these programs targeted a cohort of preterm babies and only one study assessed children at social risk excluding preterm children. In any case, and although the age of the babies, the assessment tools and the programs differed among studies, all interventions involved components of guidelines for parents to stimulate child development through improving reciprocity in parent/baby relationship and a better understanding of the child´s needs. In spite of the mentioned heterogeneities in study design, the overall conclusion was that *Early Family-Centered Intervention Programs* improved the cognitive and motor development in preterm infants when compared to the standard care.

# 4. PERSONALIZED AND PREVENTIVE MEDICINE: FROM EARLY LIFE HEALTH MONITORING TO EARLY LIFE PREVENTION PROGRAMS

Figure 1 summarizes how we envision the connection between prospective studies and the identification of biomarkers with the administration of EIP that will lead to a personalized and preventive medicine for babies exposed to PS.

In the **Prospective Study** section of the Figure we illustrate that stressed pregnant women in their third trimester are categorized as stressed with PSS (Cohen Perceived Stress Scale questionnaire, PSS-10) and controls matched for 1:1 for parity, maternal age, and gestational age at study entry. Two and a half weeks after screening, a transabdominal ECG (taECG) recording at 900 or 1000 Hz sampling rate of at least 40 min is performed to compensate for inevitable signal artifacts. From fetal and maternal ECG, fetal and maternal R-peaks are detected and to derive the FHR and MHR. The bPRSA method is used to quantify interactions between FHR and MHR as a measure of transfer of maternal stress onto fetus (Fetal Stress Index, FSI). This measure uniquely captures biophysical dynamics of mother-fetus dyad. Maternal serum is obtained to detect total cholinergic status as the total capacity for acetyl choline (ACh) hydrolysis (that is, the summation of ACh esterase, AChE, and butyl choline esterase, BChE, activities). Cholinergic status, especially the ratio of AChE/BChE, has been shown to be a sensitive indicator of chronic stress exposure in adults, but its role as stress biomarker in a developing organism remains to be fully elucidated [108-110]. AChE activity levels are assessed in the maternal serum and cord blood samples with a specific BChE inhibitor, by using a microtiter plate assay (MPA). On the day of parturition, hair strands are collected from the posterior vertex region on the head, as close to the scalp as possible, for cortisol measurement using Automated Chemiluminescent ImmunoAssay (CLIA). Based on an approximate hair growth rate of 1 cm per month, the proximal 3 cm long hair segment is assumed to reflect the integrated hormone secretion over the three-month-period prior to sampling [111].

Soon after birth, a neonatal specialized midwife collects a saliva/buccal sample from the newborn by gentle rubbing the gums on both sides with a special device and stores it at room temperature. DNA from the infant saliva sample is extracted and methylation is measured using the Infinium HumanMethylation EPIC beadchip array or similar technologies. Cord blood serum for miRNA detection and AChE/BChE analysis is also collected.

At 24 months of age, infants' development is assessed by the german version of Bayley Scale of Infant Development Third Edition (BSID-III). The BSID-III is composed from a series of subtests aimed at evaluating cognitive, language and motor skills on infants from 16 days to 42 months. A specialized professional, who is blind regarding maternal stress categorization, administers the tests, which lasts about 120 minutes. A new saliva sample is collected from the infant at this final visit for epigenetic analysis. Based on the assumption that PS imprints these phenotypic modalities permitting a mutual inference, we examine putative relationships between the measures derived from all epigenetic biomarkers, FHR analyses techniques and maternal and child cognitive assessments. One key challenge is the integration of heterogeneous datasets, such as multidimensional HRV indices, the epigenetic information, and biochemistry indices. With the entire multimodal cohort data, the machine learning tools build predictive models of the relationship between PS and the different multimodal biomarkers. With the generated model, clinicians can make predictions for the effects of PS on later neurodevelopment, e.g., by using taECG and cord-blood samples to predict BSID at 2 years of life.

In the **Personalized and Preventive Medicine** section of the Figure, we highlight the significance of this proposal stressing the importance of identifying early non-invasive pre- and postnatal biomarkers of brain programming due to intrauterine stress exposure. This will help to predict adverse postnatal brain developmental trajectories, which is a prerequisite for designing EIP. EIP improve long-term developmental trajectories and may prevent development of health abnormalities entirely. Using an unprecedented multimodal combination of epigenomic predictors and clinical health outcomes such as prenatal stress, FHR analyses in utero and epigenetic alterations in newborns promises new insights into the early causes of neuropsychological dysfunction during early childhood.

**CONCLUSION**

Stress and anxiety during pregnancy increase risk for poor child neurodevelopment. If stress persists during the nursing period, it will lead to deficient parenting interfering with the mother-infant attachment. This implies that during critical periods of brain development, i.e., pregnancy and nursing periods, the baby is subjected to environmental negative influences known to shape developmental trajectories, including neuronal connections. This apparently healthy baby, if exposed to a repeated stressful situation later in life, may show impairments in the functional development of affective and reward circuits, cognition and response inhibition.

There is now a general consensus that both stress systems, namely the HPA and the SAMS, leave permanent epigenetic marks in selected genes that if adequately linked to the neurodevelopmental disorders can emerge as reliable predictive biomarkers. In addition, simultaneous maternal-fetal heart rate monitoring is emerging as a novel early PS biomarker. Since the pregnant and nursing mother can be unaware of her stress situation or of the harmful consequences to her child, the identification and characterization of biomarkers that can timely predict adverse postnatal brain developmental trajectories, is urgently needed and a prerequisite for designing therapeutic interventions. The recently developed Early Family–Centered Intervention programs aim to support family dynamics in the domestic environment and are highly recommended due to the possibility of improving the care and stimuli offered daily to the infant.

By integrating multiple non-invasively obtainable sources of information using novel epigenetic, electrophysiological and machine learning methods, unambiguously linked to the disorders, this approach could yield progress in maternal–fetal medicine, pediatrics and developmental psychology, offering a more precise and personalized prediction and new possibilities for designing early interventions to improve neurodevelopmental outcomes of pregnancy affected by PS. This is an important first step in preventing neuropsychological problems and in reducing their personal and societal impact.

**FIGURE LEGENDS**

**FIGURE 1**

*a)* **PROSPECTIVE STUDY**

Schematic representation of a Prospective Matched Case/Control study to derive biomarkers of prenatal stress. Pregnant women in their third trimester are screened and a transabdominal electrocardiogram (taECG) is perfomed two and a half week after screening. The bivariate Phase Rectified Signal Averaging (bPRSA) method is used to quantify interactions between fetal and maternal Heart Rate (fHR and mHR) as a measure of transfer of maternal stress onto the fetus (Fetal Stress Index, FSI). Maternal serum is obtained to detect total cholinergic status as the total capacity for ACh hydrolysis by Acetyl and Butyryl Cholinesterase (AChE and BChE activities) using a microtiter plate assay (MPA). On the day of parturition maternal hair strands are collected for cortisol measurement using automated Chemoluminiscent ImmunoAssay (CLIA). Upon delivery, saliva and serum samples are collected from the newborn. DNA from the infant saliva sample is extracted and methylation is measured, e.g., using the Infinium HumanMethylation EPIC beadchip array. Cord blood serum for miRNA detection and AChE/BChE analysis, is also collected. At two years of age, the mother is invited to return with the toddler to evaluate the neurocognitive development assessed by Bayley Scale III of Infant development (BSID). Machine learning analysis will examine putative relationships between the measures derived from all epigenetic and molecular biomarkers, FSI and maternal and child cognitive assessments.

*b)* *PERSONALIZED AND PREVENTIVE MEDICINE*

Timely assessment of non-invasive biomarkers for pregnant women and newborns will allow early detection of the babies-at-risk. This will help choosing the appropiate Early Intervention Program (EIP) and improve child development. Ultimately, such approach will help prevent long-term neuropsychological problems. Such intervention programs should ameliorate future disorders in otherwise apparently healthy children reducing the personal and societal impact of such disorders.

# PROSPECTIVE STUDY

# PERSONALIZED AND PREVENTIVE MEDICINE

**SAMPLING**  **ANALYSIS & TESTS**  **OUTPUT**

## Third Trimester

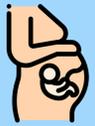

| aECG | PRSA | FSI |
| Serum | MPA | AChE/BChE |
| Hair | CLIA | Cortisol |

## Birth

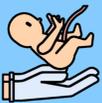

| Saliva | EPIC array | Differentially Methylated genes |
| Serum | miRNA | Differentially miRNA levels |
| Serum | MPA | AChe/BChE |

## Two Years Old

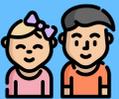

| Saliva | EPIC array | Differentially Methylated genes |
| | Bayley's Test | Combined Cognitive, Language and Motor Scores |

**MACHINE LEARNING**

**BIOMARKERS of PS**

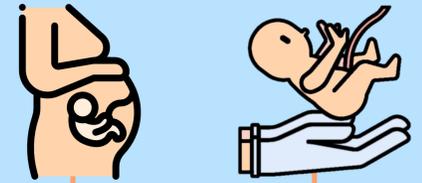

Assesment of non-invasive biomarkers

**NORMAL BABY**

**BABY AT RISK**

**NORMAL CARE**

**EIP**

**HEALTHY CHILDREN**

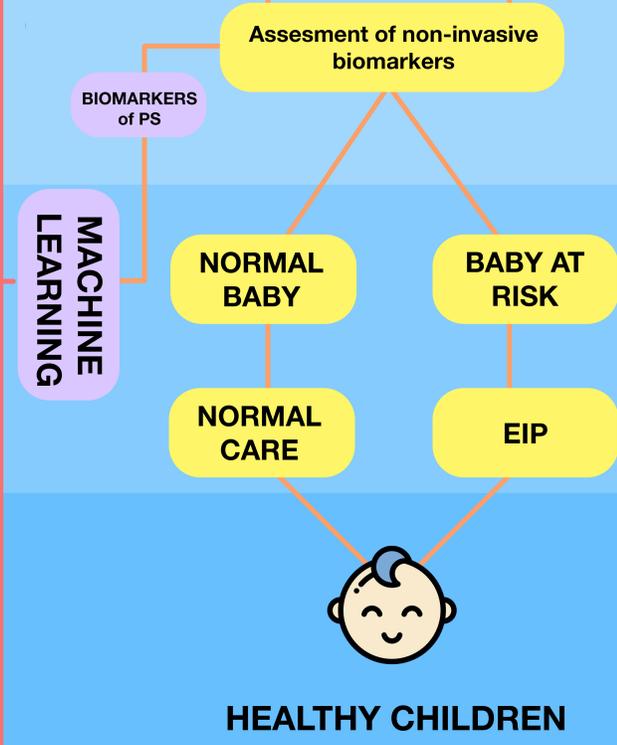